\documentclass[aps,prl,reprint]{revtex4-2}
\usepackage[normalem]{ulem}
\usepackage{amsmath,amssymb,graphicx,numprint,times,adjustbox, multirow,dcolumn,mathtools,xspace,verbatim}
\usepackage[usenames,dvipsnames,svgnames,table]{xcolor}
\definecolor{prlblue}{rgb}{0.18,0.18,0.573}
\usepackage[colorlinks=true, urlcolor=prlblue, linkcolor=prlblue, citecolor=prlblue, pdftex]{hyperref}
\usepackage{orcidlink}

\npthousandsep{\,}
\npdecimalsign{.}

\def\<{\langle}
\def\>{\rangle}

\newcommand{\prlsection}[1]{{\it #1.}---}

\newcommand{\switchtoletter}[1]{
\setcounter{equation}{0}
\renewcommand{\theHequation}{#1\arabic{equation}}
\renewcommand{\theequation}{#1\arabic{equation}}

\setcounter{figure}{0}
\renewcommand{\theHfigure}{#1\arabic{figure}}
\renewcommand{\thefigure}{#1\arabic{figure}}

\setcounter{table}{0}
\renewcommand{\theHtable}{#1.\Roman{table}}
\renewcommand{\thetable}{#1.\Roman{table}}
}
\newcounter{appcounter}
\usepackage{layouts}

\begin{document}

\title{Inducing Extraordinary-log Criticality through a  Quantum Spin Chain}
\author{\firstname{Grigorios} \surname{Makris}\,\orcidlink{0009-0005-3449-4461}}
\affiliation{\mbox{Institute for Theoretical Solid State Physics, RWTH Aachen University, 52074 Aachen, Germany}}
\author{\firstname{Francesco} \surname{Parisen Toldin}\,\orcidlink{0000-0002-1884-9067}}
\affiliation{\mbox{Institute for Theoretical Solid State Physics, RWTH Aachen University, 52074 Aachen, Germany}}
\author{\firstname{Stefan} \surname{Wessel}\,\orcidlink{0000-0002-6353-5083}}
\affiliation{\mbox{Institute for Theoretical Solid State Physics, RWTH Aachen University, 52074 Aachen, Germany}}

\begin{abstract}
We demonstrate the emergence of extraordinary-log criticality in a quantum many-body system by coupling a spin-$1/2$ Heisenberg chain to a two-dimensional quantum critical Heisenberg bilayer. Under the quantum-to-classical mapping, the chain acts as a plane defect in a three-dimensional critical environment for which the extraordinary-log universality class has recently been predicted. Using large-scale sign-problem-free quantum Monte Carlo simulations, we directly probe the renormalization-group flow of the effective defect coupling through the spin stiffness. We observe the characteristic logarithmic growth with linear system size $L$, corresponding to a stable weak-coupling fixed point and a running coupling $g(L)\sim 1/\log L$. Furthermore, the extracted step-scaling function quantitatively follows the universal renormalization-group flow predicted for the extraordinary-log universality class, including the leading and subleading universal coefficients. Our results establish extraordinary-log criticality in a genuine quantum many-body setting and show that its universal behavior persists despite the topological $\theta$-term present in the low-energy field theory of the isolated spin chain.
\end{abstract}

\maketitle
\prlsection{Introduction}
Quantum spin chains constitute a paradigmatic setting for exploring collective quantum many-body phenomena, ranging from quantum criticality and fractionalized excitations to symmetry-protected topological phases~\cite{Giamarchi2004,Chen2011,Pollmann2012,Toskovic2016}. Their relative simplicity, combined with the availability of powerful analytical, numerical, and experimental techniques, has established them as an ideal laboratory for investigating strongly correlated quantum matter. The effective low-energy description of quantum spin chains in terms of a quantum non-linear $\sigma$-model (NLSM), with classical action
\begin{equation}
    S=\frac{1}{2g}\int d^2x \: (\partial_\mu \vec n)^2, \quad |\vec{n}|=1,
    \label{nlsm}
\end{equation}
 for the order parameter field $\vec n$, complemented by a topological $\theta$-term for half-integer spin, furthermore provides a link to basic models of quantum field theory and famously rationalizes the critical vs. gapped behavior in half-integer vs. integer spin chains~\cite{haldane1983continuum, auerbach2012interacting}. 

\begin{figure}[t]
    \centering
    \includegraphics[width=\columnwidth]{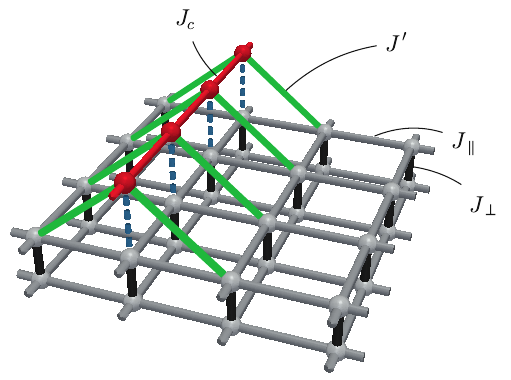}
    \caption{Two coupling geometries between a quantum Heisenberg spin chain and a quantum critical antiferromagnetic Heisenberg bilayer. In the first geometry, each spin in the chain (red) is coupled to two spins in the bulk (green bonds). In the second geometry, each chain spin is directly coupled only to a single bulk spin (via the blue dashed bonds). }
    \label{fig:setups}
\end{figure}

Here, we demonstrate that a recently predicted form of unconventional criticality — the extraordinary-log  universality class (UC) — can be realized by coupling a quantum spin chain to a quantum critical two-dimensional (2D) magnetic bulk system, see Fig.~\ref{fig:setups} for an illustration. This peculiar UC is distinguished by correlations of the order parameter   that decay extremely slowly, as a power of a logarithm of the distance. The extraordinary-log UC has been
predicted for the classical three-dimensional (3D) $O(N)$ model, at a bondary surface for $2 \leq N \leq N_c$ \cite{metlitski2022boundary} (with $N_c > 3$ \cite{parisen2021boundary, parisen2022boundary}, estimated as $N_c\approx 5$ \cite{PKMGM-21}), and on a plane defect, for any $N$ \cite{krishnan2023plane}.

The existence of this UC has previously been supported by numerical studies of $O(N)$ models for $N = 2$ and $N = 3$~\cite{parisen2025universal, parisen2021boundary,parisen2022boundary, sun2022classical, sun2022quantum, sun2023extraordinary, hu2021extraordinary}. More recently, supporting evidence has been obtained at the edge of a model for a correlated topological insulator \cite{extlog_kmh}, and
within the fuzzy-sphere approach, where the corresponding $O(N)$ boundary conformal field theory was analyzed~\cite{feng2026studying}.
Under the quantum-to-classical mapping, the present setup corresponds to a plane defect with $N=3$ for Heisenberg exchange~\cite{auerbach2012interacting}. In this work, we demonstrate the emergence of an extraordinary-log critical phase within a concrete quantum many-body model with strong quantum fluctuations.

Recent experimental advances suggest that the underlying physical setting may become accessible in engineered quantum materials: Atomically precise magnetic nanostructures and spin chains can now be assembled and probed on surfaces using scanning tunneling microscopy~\cite{Hirjibehedin2006,Loth2012,Toskovic2016}. Furthermore, the emergence of two-dimensional van der Waals magnets and magnetic heterostructures has established a versatile platform for designing low-dimensional quantum spin systems with controllable interfacial exchange couplings and proximity effects~\cite{Gibertini2019}. Most recently, highly anisotropic van der Waals antiferromagnets hosting quasi-one-dimensional spin chains have been experimentally realized, further narrowing the gap between one- and two-dimensional quantum magnetism~\cite{Park2024}. Together, these developments point towards the prospect of realizing hybrid systems in which a quantum spin chain is coupled to a two-dimensional quantum magnet, providing a natural experimental platform for the boundary quantum critical phenomena investigated in this work.

More specifically, we consider an antiferromagnetic (AF)  spin-$1/2$ Heisenberg chain with nearest-neighbor coupling $J_c$, coupled via an AF exchange $J'$ to a bilayer 2D AF Heisenberg model. The bilayer is tuned to quantum criticality by fixing the ratio of the inter-layer to intra-layer couplings, $J_{\perp}/J_{||}$  (see Fig.~\ref{fig:setups}), to 2.5220(1)~\cite{wang2006high},
realizing the 3D $O(3)$ UC. The Hamiltonian is given as 
\begin{align}
    H =J_{||} \sum_{\langle ij \rangle_{||}} \boldsymbol{S}_i&  \cdot \boldsymbol{S}_j
    + J_{\perp}\sum_{\langle ij \rangle_{\perp}} \boldsymbol{S}_i \cdot \boldsymbol{S}_j+ \notag \\
    &+ J'\sum_{\langle ij \rangle'} \boldsymbol{S}_i \cdot \boldsymbol{S}_j
    + J_{c}\sum_{\langle ij \rangle_{c}} \boldsymbol{S}_i \cdot \boldsymbol{S}_j.
\end{align}
The coupling to the spin chain corresponds to a defect for the bulk -- we show below that the critical behavior of this line defect is controlled by the extraordinary-log UC.

The defining feature of the extraordinary-log phase is that the renormalization group (RG) flow of the effective coupling $g$ in a NLSM description of the defect is given by 
\begin{equation}\label{rg equation}
    \beta(g) = -\frac{dg}{dl} = \alpha g^2 + bg^3 + O(g^4),
\end{equation}
where $\alpha$ and $b$ are universal, $N$-dependent quantities, with $\alpha>0$ for $N<N_c$ ($l$ denotes the logarithmic RG scale), rendering $g = 0$  a stable fixed point~\cite{metlitski2022boundary}. This inverted flow contrasts with the usual strong coupling limit $g\rightarrow \infty$ of the NLSM~\cite{brezin1976spontaneous, polyakov1975interaction, fateev1979quantum, berg1979computation, zinn2007phase}, and leads to the aforementioned logarithmic behavior of correlation functions and other physical properties~\cite{metlitski2022boundary, krishnan2023plane}.

From this RG equation one obtains a leading logarithmic finite-size scaling form
\begin{equation}\label{eff g}
    \frac{1}{g(L)} \approx \alpha \log L + C,
\end{equation}
where $L$ is the linear extent of the system ($C$ is not universal)~\cite{metlitski2022boundary, krishnan2023plane}. Below, we explore quantities of the quantum many-body system to  directly probe this characteristic scaling behavior of the extraordinary-log UC.

\prlsection{Setup}
For the rest of this paper, we fix the interlayer coupling to $J_{\perp} = 1$.
We consider two different setups, both of which allow for sign-problem-free quantum Monte Carlo (QMC) simulations to probe the critical properties in an unbiased way, cf. Fig.~\ref{fig:setups}.
While the direct coupling of the chain in Fig.~\ref{fig:setups} (blue dashed lines) appears more natural, its asymptotic behavior in the limit of strong $J'$ is affected by short distance physics, with a tendency towards singlet formation on the $J'$-bonds. This dimerized behavior dominates on finite-size systems, with similarities to non-critical two-leg quantum ladder systems. Still, in an intermediate regime of $J'\sim J_{\perp}$  one can robustly identify  extraordinary-log scaling \cite{SM}. In the following, we instead focus on the setup of Fig.~\ref{fig:setups} where each spin chain site is connected to two bulk spins, which avoids this asymptotic short-distance behavior. We simulated finite systems with  $2 L^2 + L$ spins and periodic boundary conditions, for $L$ up to  $176$. The QMC simulations were performed using the stochastic series expansion (SSE) method \cite{SK-91, Sandvik-99, Weber2024SSE}.  Consistent with the emergent Lorentz invariance at the (2+1)-dimensional $O(3)$ quantum critical point \cite{wang2006high}, the inverse temperature was scaled as $\beta = 2L/J_{||}$.
Note that our setup  differs from the ones studied previously~\cite{krishnan2023plane, parisen2025universal}: we introduce a defect to the critical bulk for finite $J'$, but we do not modify any couplings within the bulk itself.

Upon varying either of the two relevant AF couplings in the regime $J_c,J'>0$, we observe no indication of a phase transition at finite coupling, i.e., this regime belongs to a single phase. We illustrate this finding explicitly for the tuning of $J'$ below. While any value of $J_c$ could therefore be chosen for a quantitative analysis of this phase, additional corrections to the estimated value of $g$ may arise from a mismatch between the effective excitation velocity along the defect line and that in the bulk~\cite{metlitski2022boundary} for the finite systems accessible to QMC. We therefore fine-tune $J_c$ such that the velocity of the isolated ($J'=0$) spin chain matches that of the isolated bulk.
This choice is motivated by the empirical observation that, in the coupled system with $J'>0$, the effective velocity then remains close to the bulk velocity even for comparatively small finite systems. Based on the velocities extracted previously for the chain and the bilayer in~\cite{sen2015velocity}, we set $J_c=0.4796$. Below, we focus on this particular value of $J_c$ and present results for other values of $J_c$ in the Supplemental Material~\cite{SM}.

\begin{figure}[t]
    \centering
    \includegraphics[]
    {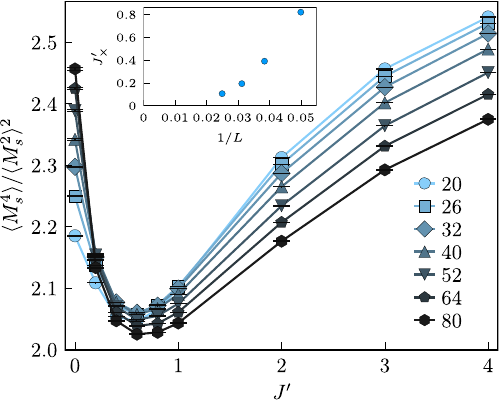}
    \caption{Binder ratio $\langle M_s^4\rangle/\langle M_s^2\rangle^2$ of the staggered magnetization of the coupled spin chain as a function of $J'$. In the inset the estimated  crossing points, $J'_{\times}$ of the Binder ratio for system sizes $L$ and $2L$ are plotted against inverse linear system size $1/L$.}
    \label{fig:binder_ratio}
\end{figure}

To  explore the possibility of a quantum phase transition upon tuning $J'$ at fixed $J_c$, we examine the finite-size scaling of the Binder ratio of the staggered magnetization, $\langle M_s^4\rangle/\langle M_s^2\rangle^2$. At a quantum critical point, this dimensionless quantity becomes independent of system size, such that curves for different $L$ cross at the critical coupling, up to finite-size corrections. The evolution of these crossings with system size therefore provides a sensitive diagnostic of both the location of a transition and the nature of the finite-size crossover.
Here, we consider the staggered magnetization of the spins on the chain ($C$),
$M_s=\sum_{i\in C}(-1)^i S^z_i$,
and plot the Binder ratio as a function of the coupling $J'$ in Fig.~\ref{fig:binder_ratio}. We find that, as $J'$ increases from zero, the finite-size trend of the Binder ratio reverses, giving rise to crossings that drift towards smaller values of $J'$. The inset shows this drift, which is consistent with a crossover scenario in which the reversal of the finite-size trend shifts towards $J'=0$ in the thermodynamic limit, and no further transition is observed at finite $J'$. Note that the decrease of the Binder ratio with increasing system size at larger values of $J'$ is also consistent with the extraordinary-log finite-size scaling prediction~\cite{parisen2025universal}.

\prlsection{Observables}
Directly identifying a logarithmic decay of correlations in finite systems is challenging, as it can easily be indistinguishable from a slow power-law.
To probe for the extraordinary-log  UC, 
we  adapt diagnostic observables, established for classical systems~\cite{parisen2025universal}, to their quantum analogs.
A central quantity in this respect is the spin stiffness $\rho_s$, quantifying the response of the coupled system to a spin torsion applied along the lateral direction. 

Indeed, the stiffness gives access to the coupling constant $g$ in the NLSM description and can be employed to define a renormalized coupling constant \cite{molke2001helicity}, allowing us to follow the RG flow.

The evaluation of $\rho_s$ in QMC 
amounts to calculating spin winding number fluctuations
~\cite{sandvik1997finite}.

The dimensionless quantity $L\rho_s$ for both the isolated critical bulk and the isolated spin chain approach constant values in the thermodynamic limit~\cite{bonvca1994direct, shastry1990twisted, wang2006high}.
Therefore, an increase of $L\rho_s$ with system size in the coupled  system ($J'>0$) showcases non-trivial collective behavior, which, as we show below, can be fitted to a logarithmic scaling form.

To  leading order in $g\rightarrow 0$, the stiffness is obtained as \cite{molke2001helicity}
\begin{equation}\label{rho L}
    L\rho_{s} \approx \frac{2v}{N}\frac{1}{g(L)} \approx \frac{2v}{N}( \alpha \log L+C)\,,
\end{equation}
where the excitation velocity $v$ emerges after the quantum-to-classical mapping,
and in the second step we employed the RG flow of $g$, Eq.~(\ref{eff g}).

The effective velocity $v$ can be independently estimated from the decay of the imaginary-time-dependent spin correlation function along the chain~\cite{SM}. Due to a finite energy gap  at  momentum $q = \pi$, corresponding to a small finite-size energy gap of the underlying critical bilayer at momentum $(\pi,\pi)$~\cite{sen2015velocity}, we extract an estimate for the velocity 
\begin{equation}\label{velocity estimation}
    v = \Delta(q)\frac{L}{2\pi},
\end{equation}
from the gap $\Delta(q)$ at $q= \pi + 2\pi/L$.
The value of $v$ is found to be around $0.8$ at $L = 80$ (see the Supplemental Material~\cite{SM} for more detailed estimations). Indeed, this is already close to the velocity of the critical bilayer estimated using the same estimator in Ref.~\cite{sen2015velocity}. In fact, there it is shown that the estimator Eq.~\eqref{velocity estimation} slightly overestimates the actual velocity, which is obtained as $v=0.75341(8) J_{\perp}$ by refined measures (that are  however not feasible for the coupled chain system). Thus, the actual effective chain velocity $v$ may also be slightly below  the above estimate. Furthermore, $v$ is expected to flow (logarithmically slowly) to the velocity of the critical bilayer bulk in the thermodynamic limit~\cite{metlitski2022boundary}. For these reasons, we use the value of $v=0.75341$ in the analysis below.

\begin{figure}[t]
    \centering
    \includegraphics
    {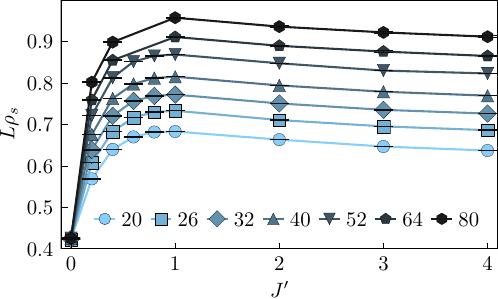}
    \caption{Spin stiffness $L\rho_s$ as a function of $J'$ for various  linear system sizes $L$.}
    \label{fig:stiff_Jcb}
\end{figure}

\prlsection{Results} 
As shown in Fig.~\ref{fig:stiff_Jcb}, the value of $L \rho_s$ increases with $L$ at any finite value of $J'$. This is in marked contrast to the behavior in the decoupled limit ($J'=0$). From Eq.~\eqref{rho L}, such a sub-linear increase is indeed anticipated in the extraordinary-log phase. Furthermore, we observe a peak at $J'\approx 1$. According to Eq.~\eqref{rho L}, in the extraordinary-log phase a higher value of $L\rho_s$ indicates a smaller value of $g$ for comparable velocities, allowing us to access the asymptotic regime $g\rightarrow 0$ on smaller lattices. 

\begin{figure}[t]
    \centering
    \includegraphics[]
    {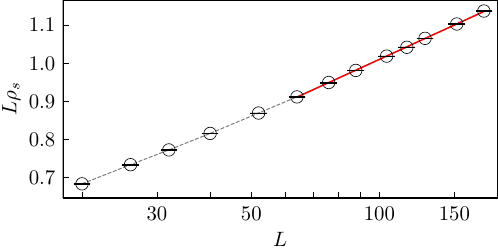}
    \caption{Spin stiffness  $L\rho_s$ as a function of $L$ for $J'=1$ on a semi-log plot (the dashed line is a guide to the eye). The logarithmic fit with respect to system size is performed for $L \geq 64$ (full line). }
    \label{fig:finite_size_stiffness}
\end{figure}

Figure~\ref{fig:finite_size_stiffness} indeed exhibits  a logarithmic growth of  $L\rho_s$ for $J'=1$ with respect to $L$, where the fit is feasible  for $L>60$. Fitting the data to Eq.~\eqref{rho L}, we obtain a value of $\alpha = 0.4412(18)$ with a quality of fit $\chi^2/\mathrm{d.o.f.}\approx 0.92$. This estimate falls below the value $\alpha = 0.540(8)$ extracted from large-scale Monte Carlo studies on classical systems~\cite{parisen2025universal}. 

The leading correction in the estimate of $\alpha$ from finite-size data stems from the next order term in the RG equation for the flow of $g$: In particular, if one takes into account the cubic term in Eq.~\eqref{rg equation}, at an intermediate scale the value of $\alpha$ effectively observed is $\alpha + b g$. Based on our estimate for $g$, and using the value of $b=-0.12(3)$ from Ref.~\cite{parisen2025universal}, this leads to the estimate $\alpha \approx 0.487(15)$ which is indeed closer to the extracted value. 
In order to obtain a refined estimate for $\alpha$ via this direct analysis, one would need to  study even larger system sizes, which is, however, computationally very demanding.

\begin{figure}[t]
    \centering
     \includegraphics[]
    {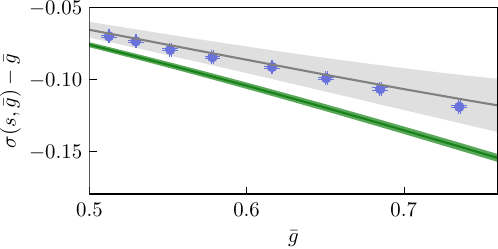}
    \caption{Step scaling function $\sigma(s,\bar g)$ for $s = 2$ as a function of $\bar g$, after subtracting the trivial linear contribution $\bar g$. The gray strip is obtained using the RG flow of Eq.~(\ref{rg equation}) with  the universal constants, including their uncertainties, taken from the classical simulations in Ref.~\cite{parisen2022boundary, parisen2025universal}, $\alpha = 0.540(8)$ and $b = -0.12(3)$. The green strip corresponds to truncating the RG flow at quadratic order (i.e., setting $b=0$). Both strips tend towards $0$ and approach  each other as $g$ approaches 0 (not shown). The blue points show the values of $\sigma(s,\bar g) - \bar{g}$ obtained from our simulations for the coupled quantum spin chain system.}
    \label{fig:step_scaling}
\end{figure}

Nevertheless, there is a more practical way to verify that the flow of $g$ is controlled  by Eq.~\eqref{rg equation} with the universal coefficients predicted in Ref.~\cite{parisen2022boundary, parisen2025universal}, based on feasible QMC simulations. To this end, we consider the step scaling function~\cite{luscher1991numerical} 
\begin{equation}\label{step scaling}
    \sigma(s,\bar g(L)) := \bar g(sL),
\end{equation}
where $\bar{g}(L)\equiv 2v/(L\rho_sN)$ is a renormalized coupling constant defined on the basis of Eq.~\eqref{rho L}, with $\bar g = g + O(g^2)$, and  $s$ the  step parameter that we fix to 2. Based on the RG flow  in Eq.~(\ref{rg equation}) one infers the expansion 
\begin{equation}\label{sigmaexp}
\sigma(s,\bar g)=\bar g-(\alpha \ln s) \bar g^2 +[\alpha^2 (\ln s)^2-b \ln s]\bar g^3+O(\bar g^4)
\end{equation}
for the step scaling function~\cite{luscher1991numerical}.
As shown in Fig.~\ref{fig:step_scaling},
our  QMC data for
$\bar{g}(L)$ indeed fall onto the universal curve for $\sigma(s,\bar{g})$  given by Eq.~(\ref{sigmaexp}),
with the universal values for $\alpha$ and $b$ again taken from Ref.~\cite{parisen2022boundary, parisen2025universal}. This demonstrates the emergence of the characteristic RG flow underlying the extraordinary-log UC in the coupled quantum spin chain system on a  quantitative level -- based on the accessible system sizes for unbiased QMC simulations. Fig.~\ref{fig:step_scaling} furthermore demonstrates the relevance of the cubic term in the RG flow on accessible system sizes, for which the finite-size estimate $\bar g(L)\sim O (1)$ is beyond the asymptotic scaling regime, where $g\rightarrow 0 $.

\prlsection{Conclusions}
Based on large-scale QMC simulations, we provided strong  evidence for the emergence of 
extraordinary-log criticality in a  spin-1/2 Heisenberg chain coupled to a quantum critical two-dimensional bulk system. For this purpose, we directly targeted the RG flow of the effective coupling $g$, which is accessible via the system's spin stiffness, exhibiting  logarithmic scaling with system size -- characteristic of the extraordinary-log UC. Taking both the leading and subleading contribution to the RG flow of $g$ in Eq.~\eqref{rg equation} into account, in terms of the universal coefficients $\alpha$ and $b$~\cite{parisen2022boundary, parisen2025universal}, we observed quantitative agreement with finite-size data 

-- also demonstrating that even on 
comparably large systems, the RG flow of $g$ is controlled by the presence of both  universal  contributions.

It is worthwhile to note that we observe the extraordinary-log scaling even though the effective low-energy field theory of the spin-1/2 chain contains a topological $\theta$-term~\cite{haldane1983continuum}, which is absent in the classical NLSM description of the extraordinary-log UC. This indicates that the $\theta$-term is apparently irrelevant for the extraordinary-log UC in the quantum many-body setting. This possibility was noted previously~\cite{metlitski2022boundary}, arguing that  the  perturbation theory in $g$, from which the RG flow in Eq.~\eqref{rg equation} derives, is not modified by the topological term.

Moreover, it is well known that the $\theta$-term vanishes for integer-spin Heisenberg chains~\cite{haldane1983continuum,haldane1983nonlinear, auerbach2012interacting}. Therefore, studying a setup in which a spin-1 chain is coupled to the critical bilayer  provides an alternative possibility to induce extraordinary-log scaling. Since the spin-1 chain is gapped however, the anticipated gap closing may then be accompanied by a boundary phase transition at a {\it finite} coupling between the chain and the bilayer. For the future, it will thus be interesting to examine this setup, which may in addition to the extraordinary-log UC showcase the ordinary UC \cite{Diehl-86,PT-23} and the special surface transition \cite{parisen2021boundary} as well in  a single quantum many-body setting.

\begin{acknowledgments}
F.P.T. thanks Max Metlitski for useful discussions.
F.P.T. is funded by the Deutsche Forschungsgemeinschaft (DFG, German Research Foundation), Project No. 586743423. The authors would like to thank the German Federal Ministry of Research, Technology and
Space and the German federal states (https://go-nhr.de/our-partners) for supporting this
work/project as part of the National High-Performance Computing (NHR) joint funding
program. The QMC simulations were performed using a locally modified version of the StochasticSeriesExpansion.jl package~\cite{Weber2024SSE}, and the data post-processing was performed based on  the Carlo.jl package~\cite{weber2025carlo}. 
All the numerical data shown in the study is available online~\cite{Makris2026}.

\end{acknowledgments}

\bibliographystyle{apsrev4-2}
\bibliography{refs}

\clearpage

\onecolumngrid
  \parbox[c][3em][t]{\textwidth}{\centering \large\bf Supplemental Material}
\smallskip
\twocolumngrid

\prlsection{Velocity estimation}
Within the extraordinary-log phase, the spin excitation velocity of the chain/defect is expected to flow to the velocity of the bulk in the thermodynamic limit \cite{metlitski2022boundary}. We examine the decay of imaginary time displaced correlation functions $\langle S^z_q(\tau)S^z_{-q}(0)\rangle$ and extract the velocity according to Eq.~\eqref{velocity estimation} of the main text, fitting the data to a single decaying exponential for $\tau$ within the range $ [0.14,0.32]\beta$.  Three different supports for $S_q^z$ have been considered: For the first quantity, $S_{q,c}^z$, only the spins on the chain are considered. For the second, $S_{q,d}^z$, only the spins of the bulk to which the chain is coupled are considered. Lastly, we estimate the spin excitation velocity of the total system based on all spins, $S_{q,b}^z$. For linear system sizes $L = 20,40,80$ we obtain the corresponding velocity estimates $v_c = 0.8207(19), 0.812(5) , 0.7946(56)$, $v_d = 0.8185(18),0.810(3) , 0.798(6)$  and $v_b = 0.7928(9), 0.7925(16), 0.782(7)$, respectively. We observe that for the the mean values of $v_c$ and $v_d$ agree with the value $v_b$ to order $3.5\%$. This strongly supports our use of $v$ from Ref.~\cite{sen2015velocity} throughout this work, corresponding to a refined estimation of the bulk velocity compared to the one obtained from Eq.~\eqref{velocity estimation} of the main text.

\prlsection{Adapted stiffness estimator for large $J_c$}
Since the spin stiffness of an isolated spin chain  increases with $J_c$~\cite{bonvca1994direct, shastry1990twisted}, we expect that for large  $J_c$  the total spin stiffness of the coupled system will be dominated by winding number fluctuations within the chain, substantially affecting the estimation of the parameter $\alpha$ by the spin stiffness via Eq.~(\ref{rho L}). 

In order to correct for this bias, we have to disentangle the direct contribution of the large chain winding  number fluctuations to the total stiffness. Here, we introduce for this purpose a modified winding number estimator, in which the component that is correlated with the chain winding number is projected out: Namely, we can decompose the total winding number  as
\begin{equation}
\label{W_tot}
W_{\mathrm{tot}}=W_b+W_c,
\end{equation}
where $W_b=N_+^b-N_-^b$ and $W_c=N_+^c-N_-^c$ are the winding numbers measured in the bulk and on the chain, respectively. The usual stiffness estimator is proportional to
\begin{equation}
\label{Wsquare_tot}
\left\langle W_{\mathrm{tot}}^2\right\rangle
=
\left\langle W_b^2\right\rangle
+
\left\langle W_c^2\right\rangle
+
2\left\langle W_bW_c\right\rangle.
\end{equation}

We define an effective winding number
\begin{equation}
\label{variational_form}
W_{\mathrm{eff}}=W_{\mathrm{tot}}-\lambda W_c
\end{equation}
and choose $\lambda$ such that $W_{\mathrm{eff}}$ is not correlated with the chain winding number,
\begin{equation}
\label{definition}
\left\langle W_{\mathrm{eff}}W_c\right\rangle=0.
\end{equation}

\begin{figure}[!t]
    \centering
    \includegraphics[]
    {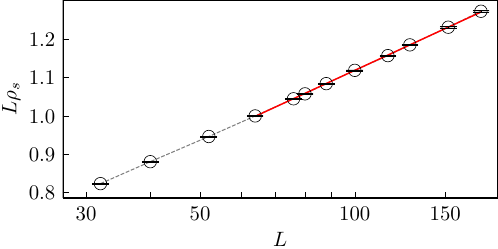}
    \caption{Spin stiffness  $L\rho_s$, obtained using the effective winding number estimator, Eq.~\eqref{effective_estimator}, for the doubly connected chain at $J'=1$ and $J_c = 4$ on a semi-log plot (the dashed line is  a guide to the eye). The logarithmic fit is shown by the full line.}
    \label{fig:stiffness_Jc4_effective}
\end{figure}

Since the mean winding numbers vanish in our simulations, this condition is equivalent to requiring zero covariance. We obtain
\begin{equation}
\lambda
=
\frac{\left\langle W_{\mathrm{tot}}W_c\right\rangle}
{\left\langle W_c^2\right\rangle}
=
1+
\frac{\left\langle W_bW_c\right\rangle}
{\left\langle W_c^2\right\rangle}.
\end{equation}
Consequently,
\begin{equation}\label{effective_winding}
W_{\mathrm{eff}}
=
W_b-
\frac{\left\langle W_bW_c\right\rangle}
{\left\langle W_c^2\right\rangle}W_c.
\end{equation}

Its fluctuations are therefore given by
\begin{equation}
\label{effective_estimator}
\left\langle W_{\mathrm{eff}}^2\right\rangle
=
\left\langle W_b^2\right\rangle
-
\frac{\left\langle W_bW_c\right\rangle^2}
{\left\langle W_c^2\right\rangle}.
\end{equation}

The same value of $\lambda$ minimizes
\begin{equation}
\left\langle
\left(W_{\mathrm{tot}}-\lambda W_c\right)^2
\right\rangle,
\end{equation}
so $W_{\mathrm{eff}}$ can also be interpreted as the residual winding number after subtracting the component of $W_{\mathrm{tot}}$ that is linearly correlated with $W_c$. 

It is important that this construction does not imply $ \left\langle W_{\mathrm{tot}}^2\right\rangle
=
\left\langle W_{\mathrm{eff}}^2\right\rangle
+
\left\langle W_c^2\right\rangle $.
Instead, defining
\begin{equation}
W_c'
=
\left(
1+
\frac{\left\langle W_bW_c\right\rangle}
{\left\langle W_c^2\right\rangle}
\right)W_c,
\end{equation}
one has
\begin{equation}
W_{\mathrm{tot}}=W_{\mathrm{eff}}+W_c',
\qquad
\left\langle W_{\mathrm{eff}}W_c'\right\rangle=0,
\end{equation}
and hence
\begin{equation}
\left\langle W_{\mathrm{tot}}^2\right\rangle
=
\left\langle W_{\mathrm{eff}}^2\right\rangle
+
\left\langle {W_c'}^2\right\rangle.
\end{equation}

\begin{figure}[t]
    \centering
     \includegraphics[]
    {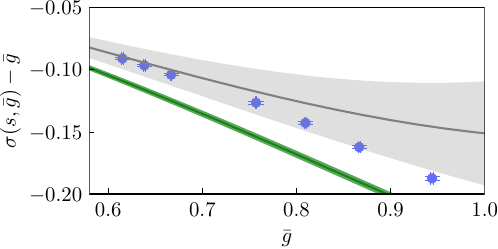}
    \caption{Step scaling function $\sigma(s,\bar g)$ for $s = 2$ as a function of $\bar g$, for the setup of the doubly connected case at $J_c = 4, J' = 1$ and lattice sizes up to $L=88$, with estimations $\bar{g}$ obtained through the stiffness using the effective estimator $\langle W_\mathrm{eff}^2\rangle$ of Eq.~\eqref{effective_estimator}. The gray and green ribbons are obtained as detailed in the  descriptions of Fig.~\ref{fig:step_scaling}. }
    \label{fig:step_scaling_Jc4_effective}
\end{figure}

\begin{figure}[t]
    \centering
    \includegraphics[]{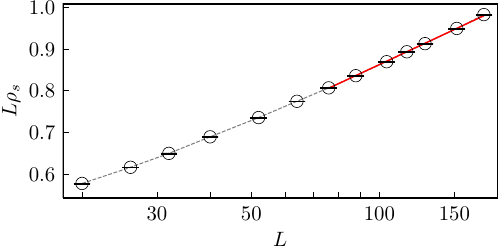}
    \caption{  Spin stiffness  $L\rho_s$ as a function of $L$ for $J'=1$ for the singly connected chain setup on a semi-log plot (the dashed line is  a guide to the eye). The logarithmic fit is shown by the full line.  }
    \label{fig:single_stiffness}
\end{figure}

\begin{figure}[t]
    \centering
     \includegraphics[]
    {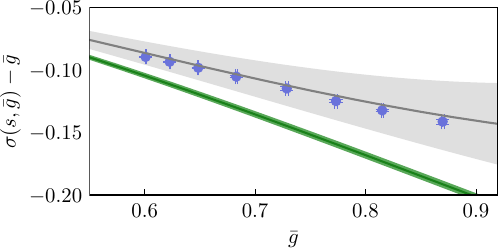}
    \caption{Step scaling function $\sigma(s,\bar g)$ for $s = 2$ as a function of $\bar g$, for the setup of the singly connected chain at $J_c = 0.4796, J' = 1$ and lattice sizes up to $L=88$. The gray and green ribbons correspond to the descriptions of Fig.~\ref{fig:step_scaling}. }
    \label{fig:step_scaling_single}
\end{figure}

This construction projects out the component of the winding number that is parallel to $W_c$. It does not ensure that the full microscopic contribution of the chain has been removed, nor that $\langle W_{\mathrm{eff}}^2\rangle$ is by itself a conventional residual stiffness after removing the chain contribution. We therefore use it merely as a diagnostic estimator at large $J_c$, where the conventional spin stiffness estimator $\rho_s=\langle W_{\mathrm{tot}}^2\rangle/(\beta L^2)$ is affected by enhanced winding fluctuations associated with the strong intra-chain coupling. Using instead $\langle W_{\mathrm{eff}}^2\rangle$ in the stiffness estimator and the estimated effective velocity for  this value of $J_c$,  $v \approx 1$, we obtain a finite-size scaling fit with $\alpha = 0.4026(20)$ (with $\chi^2/\mathrm{d.o.f.} \approx 0.77$), cf. Fig.~\ref{fig:stiffness_Jc4_effective}, and a step-scaling function plot,  
Fig.~\ref{fig:step_scaling_Jc4_effective}, which is reasonably compatible with the RG flow in Eq.~\eqref{rg equation}.
We finally note that the above value of $\alpha$ is close to that obtained from the fit for the singly connected chain, Fig.~\ref{fig:single_stiffness}, where $\alpha = 0.4125(22)$. This is noteworthy and probably not accidental since it shows non-trivial consistency; in both cases the estimated values of $\bar{g}$ indeed  fall within the same range, as shown by Figs.~\ref{fig:step_scaling_single} and~\ref{fig:step_scaling_Jc4_effective}, respectively.

\prlsection{Single connection between chain and bulk}
For the case of a single connection (i.e., through a single coupling per spin) between the chain and the bilayer, we again examined both the 
 finite-size scaling of the spin stiffness as well as the step scaling function fit. The value of $\alpha$ extracted from the fit of the spin stiffness in Fig.~\ref{fig:single_stiffness} according to Eq.~\eqref{rho L}, using the velocity $v$ of Ref.~\cite{sen2015velocity}, is $\alpha = 0.4125(22)$. The quality of the fit is $\chi^2/\mathrm{d.o.f.} \approx 2.6$; it changes to $\chi^2/\mathrm{d.o.f.} \approx 1.8$ if the fit starts from  $L = 88$, while the estimation of $\alpha$ varies within the uncertainty.
The difference in the estimate of $\alpha$ from the one in the main text can be attributed to the fact that the corresponding values of $g$ accessed in this coupling setup are higher. To more accurately examine whether extraordinary-log scaling is applicable, we also consider the step-scaling function plot, cf. Fig.~\ref{fig:step_scaling_single}. We again observe compatibility with the universal RG flow of Eq.~\eqref{rg equation}, using the estimates for the universal constants from Ref.~\cite{parisen2022boundary, parisen2025universal}.

\begin{figure}[tb]
    \centering
    \includegraphics[]
    {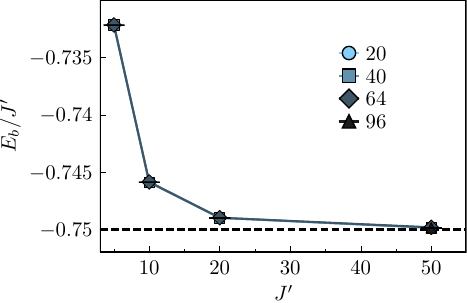}
    \caption{Bond energy $E_b=J'\langle \mathbf{S}_i\cdot \mathbf{S}_j\rangle$ along a single $J'$-bond  as a function of  $J'$  for the singly connected chain setup with $J_c=0.4796$, and for different system sizes $L$ (the case of $L = 96$ was only simulated at $J' = 50$). The dashed line indicates the bond energy $E_b=-3/4 \: J'$ for an exact  $J'$-dimer singlet ground state. 
 }
    \label{fig:bond_energy}
\end{figure}

Furthermore, as discussed in the main text, the case of a single coupled chain is  subtle, since in the large-$J'$ limit  the formation of dimer singlets between the chain spins and the bulk spins to which they are directly coupled becomes important. In fact, as shown in Fig.~\ref{fig:bond_energy}, with increasing $J'$ we 
observe a strong trend towards the asymptotic dimerized state. 

\begin{figure}[tb]
    \centering
    \includegraphics[]
    {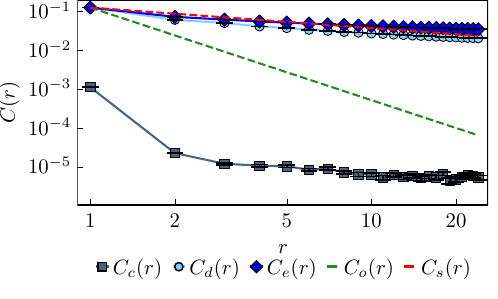}
    \caption{Staggered spin correlation functions for distances up to $r=96/4=24$ on a log-log plot. $C_{c}(r)$ gives the correlations along the chain for $J'=50$ and $L=96$ in the singly connected chain setup, and $C_{d}(r)$ the correlations among the bulk spins located directly below the coupled sites in the same geometry. $C_{o}(r)$ and $C_{s}(r)$ show  power-law decays associated with the ordinary and special UCs, respectively. $C_{e}(r)$ shows the chain correlations for the doubly connected setup studied in the main text, with $L=176$ and $J_c=0.4796$. All correlations except $C_{c}(r)$ are rescaled to the same value at $r=1$, facilitating a direct comparison of their spatial decays.}
    \label{fig:c(r)}
\end{figure}

To probe the residual extended correlations in this regime, we compare in Fig.~\ref{fig:c(r)} the spatial decay of the staggered spin correlations along the chain, $C_c(r)$, with that of the corresponding array of bulk spins located directly beneath the sites coupled to the chain, $C_d(r)$. Here, we consider the case of $J'=50$.
In the same figure, we also show  for comparison power-law decays determined by the scaling dimensions associated with the ordinary UC, $C_{o}(r)$, and the special-point UC, $C_{s}(r)$~\cite{PT-23, parisen2021boundary}.  We also include the spin correlation function along the chain for the case studied in the main text, $C_{e}(r)$, belonging to the extraordinary-log UC.

The slowest decays are observed for $C_d(r)$ and $C_e(r)$.  In particular, both exhibit slower decay than the reference lines for $C_{o}(r)$ and $C_{s}(r)$. This suggests that like $C_e(r)$,  $C_d(r)$ is also compatible with extraordinary-log behavior. While the overall magnitude of $C_c(r)$ is substantially weaker, it decays rather slowly on  larger distances, in particular, $C_c(r)$ does not exhibit an exponential decay. Instead, the data  are also compatible with extraordinary-log behavior, i.e., the spin correlations along the chain  exhibit a slow, logarithmic decay, but with a weaker overall signal at large $J'$, due to the competing dimerization in the $J'\rightarrow \infty$ limit, observed in Fig~\ref{fig:bond_energy}.

\end{document}